# Chemomechanical Origin of Hydrogen Trapping at Grain Boundaries in FCC Metals

Xiao Zhou[a], Daniel Marchand[a], David L. McDowell[b], Ting Zhu[b], Jun Song[a, *a]

[a]Department of Mining and Materials Engineering, McGill University, Montréal, Québec H3A 0C5, Canada

[b]Woodruff School of Mechanical Engineering, Georgia Institute of Technology, Atlanta, Georgia 30332, USA

**Hydrogen embrittlement of metals is widely observed, but its atomistic origins remain little understood and much debated. Combining a unique identification of interstitial sites through polyhedral tessellation and first-principles calculations, we study hydrogen adsorption at grain boundaries in a variety of face-centered cubic metals of Ni, Cu, $\gamma$-Fe and Pd. We discover the chemomechanical origin of variation of adsorption energetics for interstitial hydrogen at grain boundaries. A general chemomechanical formula is established to provide accurate assessments of hydrogen trapping and segregation energetics at grain boundaries, and it also offers direct explanations for certain experimental observations. The present study deepens our mechanistic understanding of the role of grain boundaries in hydrogen embrittlement, and promises a viable path towards predictive microstructure engineering against hydrogen embrittlement in structural metals.**

---

*[a] Author to whom correspondence should be addressed. E-Mail: jun.song2@mcgill.ca

Despite dramatic advances of polymers and composites in the past several decades, metals remain irreplaceable in many important applications for automotive, aerospace and energy industries. However, metals are typically susceptible to environmental attack. One prominent example is hydrogen embrittlement (HE) that can result in sudden and catastrophic failure of metallic components and systems (*1*). Hydrogen is abundant in service environments and manufacturing processes. As a result, HE poses a significant threat to load-bearing metallic components and is often considered as a major obstacle to the reliable applications of structural metals. Despite considerable effort in the study of HE (*2-15*), the dominant physical mechanisms of HE remain controversial (*16, 17*). Hence, the study of HE at the atomistic and electronic levels may illuminate the mechanistic origin of HE and thus enable the development of effective means to mitigate HE. Moreover, the influence of hydrogran on dislocation migration can be distinguished from its role in GB segregation and compromise of fracture resistance. We focus on the latter here.

Hydrogen adsorption is favored more at microstructural heterogeneities (*18*), such as grain boundaries (GBs), than in interstitial sites in the bulk lattice. Recently, Bechtle et al. (*19*) conducted experiments on GB-engineered Ni samples with and without hydrogen, and their results showed that the susceptibility of HE can be drastically redued at special GBs that are characterized with low excess free volumes and a high degree of atomic matching. In addition, Oudriss et al. (*20*) showed that special GBs can trap hydrogen and reduce hydrogen diffusion. These studies highlight the important role of GBs in influencing hydrogen transport and embrittlement behaviors, and suggest the possibility of controlling the susceptibility of structural metals to HE through GB engineering.

To advance rational GB engineering, it is essential to characterize the GBs and associated hydrogen segregation behaviors in a systematic and quantitative manner. The structure of certain high angle tilt GBs is commonly described by the Coincidence Site Lattice (CSL) model (*21*). Alternatively, the GB structure can be represented by a periodic array of nested three-dimensional (3D) structural units (*22, 23*), which are associated with CSL boundaries but also pertain to general high angle tilt grain boundaries. Along this line of approach, Ashby et al. (*24*) showed that there exist eight unique convex polyhedrons with triangle faces (i.e., so-called deltahedra) to account for all possible basic packing units at a tilt GB. This approach enables the characterization of GBs with a simple, yet powerful, concept of geometric packing unit, which can be applied to investigate many structural and chemomechanical properties of GBs, such as interstitial impurity segregation at GBs (*25, 26*).

Here we develop a novel modeling approach that combines the space tessellation of polyhedral packing units and first principles density functional theory (DFT) calculations for studying the hydrogen segregation at GBs in face-centered cubic (FCC) metals, including Ni, Cu, $\gamma$-Fe and Pd. The polyhedral packing units at GBs are uniquely identified and their central holes are shown to serve as favorable interstitial sites of hydrogen adsorption. Our DFT calculations reveal a universal dependence of hydrogen adsorption energies on the local volume deformation of polyhedral packing units in all four FCC metals studied. To uncover its physical origin, we establish a general formula involving a minimum number of first principles input and fitting parameters that nicely matches all DFT data of hydrogen adsorption energies at GBs in four different FCC metals. Such a general result illuminates the chemomechanical origin of hydrogen segregation at GBs. The physical meaning of the parameters in the formula is clarified to facilitate determination without fitting. Our results thus provide mechanistic insights towards predictive GB engineering to support the development of HE-resistant metals.

We have studied a number of symmetric tilt GBs with various misorientations in the FCC metals of Ni, Cu, $\gamma$-Fe and Pd. Among these GBs, five types of polyhedrons are involved, namely tetrahedron (TET), octahedron (OCT), pentagonal bipyramid (PBP), cap trigonal prism (CTP) and bi-tetrahedron (BTE) (*27*), as illustrated in Fig. 1 for a representative $\Sigma 5(130)[100]$ GB (meaning the (130) GB face with a [100] tilt axis) in Ni. We identified these polyhedrons by space tessellation (*28*). Energy minimization from DFT calculations indicates that there is only one interstitial site of hydrogen adsorption in each polyhedron, which lies closely to the centroid of the polyhedron. This is consistent with the Switendick criterion based on the minimum H-H distance (*29*). Hence, the center of each polyhedron corresponds to an individual hydrogen adsorption site at the GBs. In other words, one can identify potential hydrogen adsorption sites along GBs using a geometric approach of space tessellation of polyhedral packing units without a detailed knowledge of hydrogen adsorption chemistry.

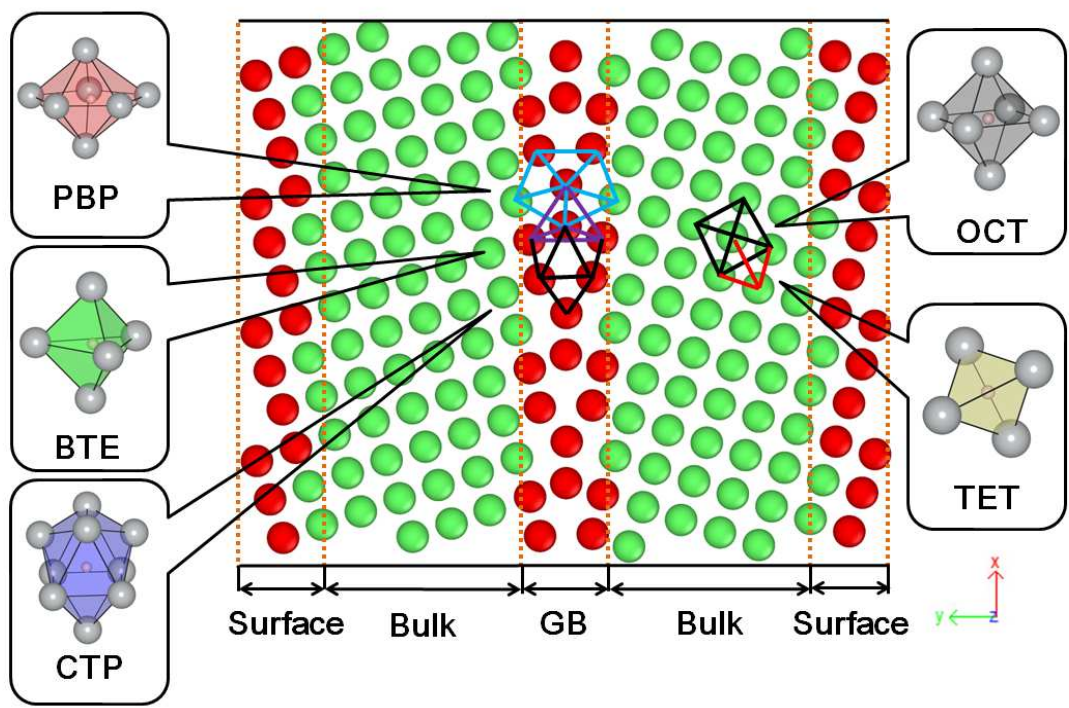

**Figure. 1** (color online): Schematic illustration of polyhedron in representative Σ5(130)[100] GB and bulk lattice. The grey ball in distinguished polyhedron represents host Ni atom and the small pink ball in center is hydrogen atom.

After the identification of hydrogen adsorption sites at GBs through space tessellation of polyhedral packing units, we performed DFT calculations to evaluate the interactions between hydrogen and GBs in terms of adsorption energetics. Fig. 2 shows the differential charge density of Σ5(130)[100] Ni GB projected along the (100) plane (see Fig. S3 for similar plots of other FCC metals studied). These results indicate that the interactions between hydrogen and the host atoms are dominantly localized at GBs, and hence the adsorption energy can be primarily determined by the local environment of the *capsule*, i.e., the polyhedron enclosing the hydrogen atom.

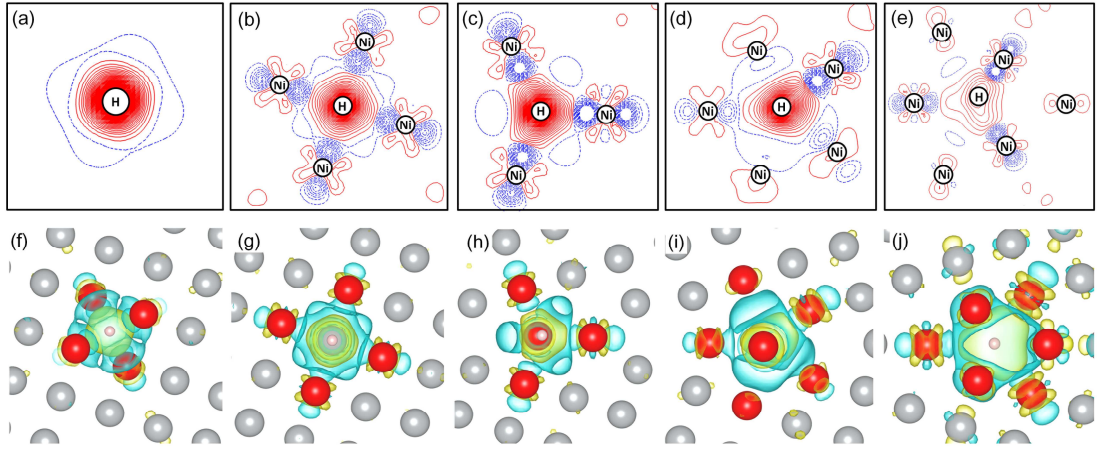

**Figure. 2** (color online): Differential charge density of $\Sigma 5(130)[100]$ Ni GB projected on the (100) plane, with (a)-(e) being two dimensional while (f)-(j) being three dimensional charge density contours. In (a)-(e), the blue dash line represents electron depletion and the red solid line signifies electron accumulation. In (f)-(j), the red spheres represent Ni atoms locating at the vertices of the polyhedron that encloses the hydrogen atom, indicated by the small pink sphere, while the grey spheres represent other Ni atoms. Blue region represents electron depletion while yellow region signifies electron accumulation.

The adsorption energy of hydrogen, $E^{ad}$, is defined as

$$E^{ad} = E_H^{GB} - E^{GB} - E_{iso}^H, \qquad (1)$$

where $E^{GB}$ and $E_H^{GB}$ are the total energies of the system before and after adsorption of one hydrogen atom, respectively, and $E_{iso}^H$ is the energy of one isolated hydrogen atom in vacuum. Based on Eq. (1), we calculated the hydrogen adsorption energies for different polyhedral sites for a series of symmetric tilt GBs (*28*) in Ni, Cu, $\gamma$-Fe and Pd, as plotted in Fig. 3. Clearly, the hydrogen adsorption energies $E^{ad}$ depend largely on the type of polyhedral interstitial site, i.e., the average values of $E^{ad}$ differ for different types of polyhedra. Moreover, for each type of polyhedral interstitial site, the values of $E^{ad}$ vary markedly. Hence, the polyhedron is not sufficient alone to uniquely determine the interactions between hydrogen and GBs.

To understand the large variation of $E^{ad}$, we examined the local deformation of polyhedrons at GBs. A parameter, $dV_p/V_p^0$, is used to measure the local volume changes (dilatation) of polyhedrons. Here, $V_p^0$ is the volume of the pristine polyhedron, which is defined as the corresponding deltahedron with the edge length being the nearest-neighboring distance, $\sqrt{2}a_0/2$, where $a_0$ denotes the equilibrium lattice constant in the bulk FCC lattice; further, $dV_p = V_p - V_p^0$ measures the deviation of the actual polyhedron volume $V_p$ from $V_p^0$. We plot $dV_p/V_p^0$ together with $E^{ad}$ in Fig. 3, and observe strong correlation between the DFT data of $E^{ad}$ and $dV_p/V_p^0$ for all four FCC metals studied.

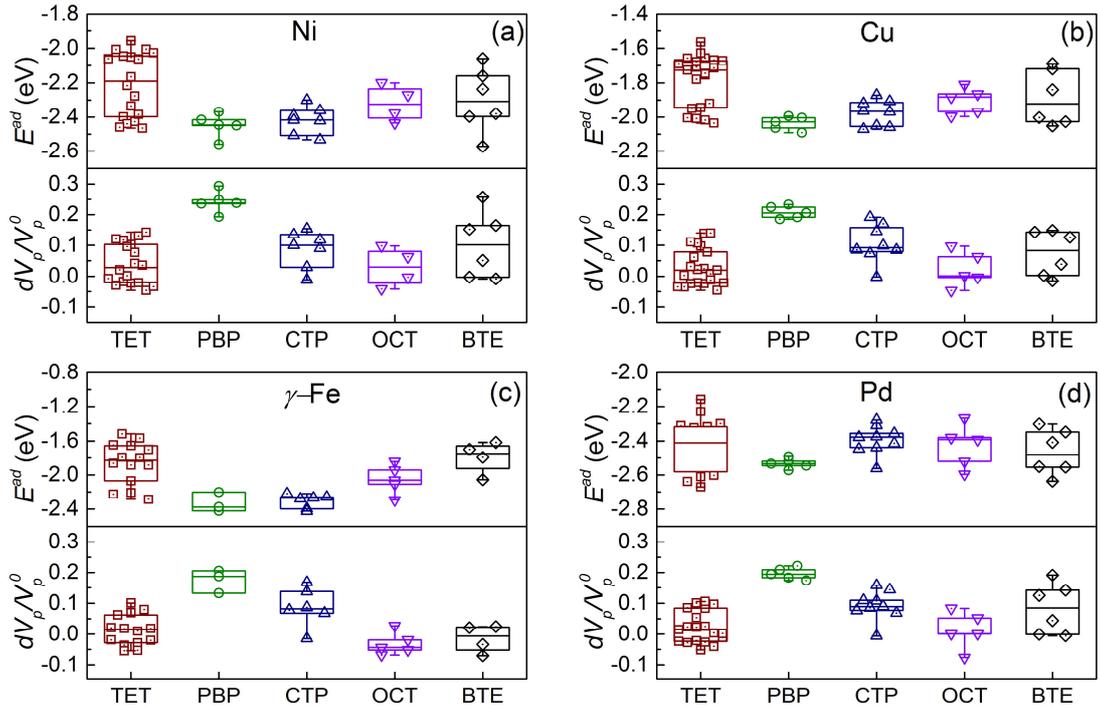

**Figure. 3** (color online): The variation in the hydrogen adsorption energy $E^{ad}$ and normalized lattice dilatation $dV_p/V_p^0$ of polyhedrons in (a) Ni, (b) Cu, (c) $\gamma$-Fe, and (d) Pd systems.

The intriguing correlation shown in Fig. 3 suggests that the variation of $E^{ad}$ is dictated by the local volume changes of polyhedral packing units. To elucidate the physical origin of such a correlation, we note that the mechanical interaction energy between a GB and an interstitial point defect can be determined by evaluating the work of $P\Omega_p$, which corresponds to the local pressure $P$ in the absence of point defects times the lattice expansion (i.e., partial volume) $\Omega_p$ occurring due to interstitial insertion of a point defect (*10, 30*). For the present case of hydrogen adsorption at the polyhedral interstitial site, the partial volume $\Omega_p$ should be the volume change associated with hydrogen adsorption at the polyhedral interstitial site, and the local pressure $P$ is determined by the volume change $dV_p/V_p^0$ according to

$$P = -B\frac{dV_p}{V_p^0} \tag{2}$$

where $B$ is the bulk modulus (see Table I). We note that the bulk modulus of the lattice is an approximation of the local bulk modulus pertaining to GB regions, owing to the differences in atomic coordination and bond lengths, but is considered here to serve as an approximation. Consequently the mechanical interaction energy between the hydrogen and polyhedral packing units is estimated as

$$dE^{ad} = -B\Omega_p \frac{dV_p}{V_p^0}. \tag{3}$$

Using Eq. (3), we can express the adsorption energy of a hydrogen atom in a polyhedron packing unit at a GB in terms of

$$E^{ad} = E_0^{ad} - B\Omega_p \frac{dV_p}{V_p^0}, \tag{4}$$

where $E_0^{ad}$ is the chemisorption energy of hydrogen in a deltahedron (i.e., a pristine polyhedron as defined earlier). Eq. (4) explicitly reveals the dependence of hydrogen adsorption on the chemisorption energy and the mechanical interaction energy, the latter of which is governed by the partial volume of hydrogen insertion and the local volume deformation of the polyhedral structural unit at GBs. In Fig. 4, the fitting curves (dashed lines) based on Eq. (4) agrees very well with the data points of adsorption energies from DFT calculations. Here for simplicity yet without loss of generality, we take a single value of partial volume $\Omega_p$ for all polyhedrons, given an FCC metal studied (*31*). The fitting parameters of $E_0^{ad}$ and $\Omega_p$ are listed in Table I (*28*).

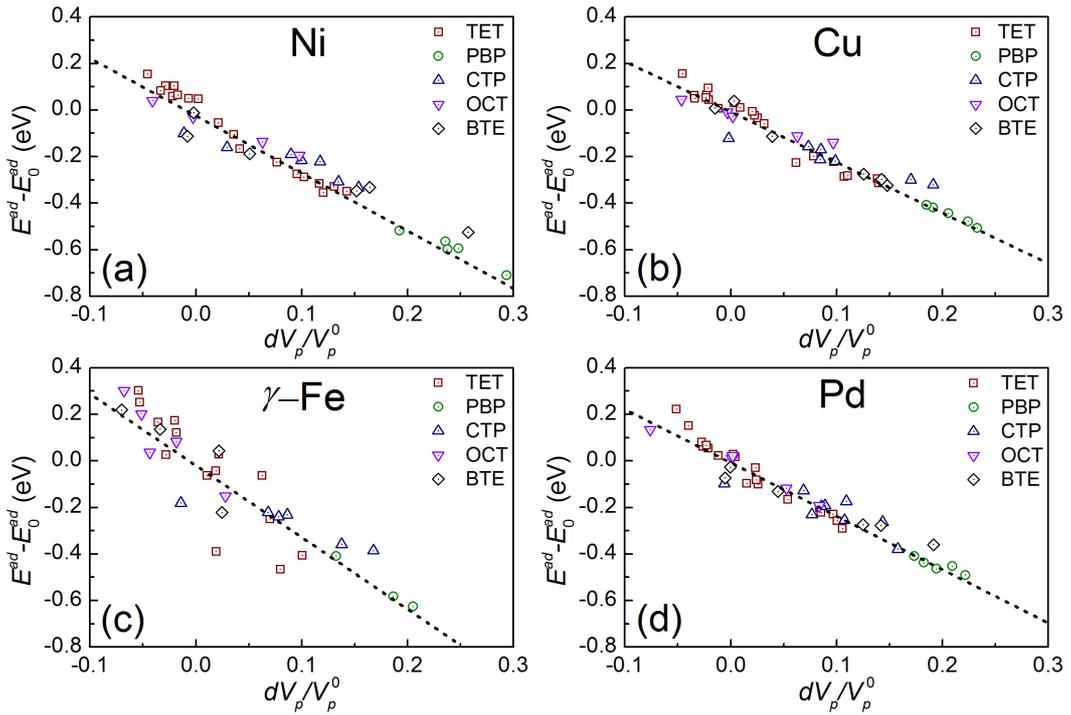

**Figure. 4** (color online): Comparison between the DFT calculated (open symbols) and model predicted (dash lines) adsorption energetics versus volume distortion relation in (a) Ni, (b) Cu, (c) $\gamma$-Fe, and (d) Pd systems.

**Table I**: List of material properties, i.e., bulk modulus (B), bulk hydrogen partial volume (Ω), predicted hydrogen partial volume at polyhedrons ($\Omega_p$), model predicted (and DFT calculated)

adsorption energy of hydrogen in a pristine polyhedron [Model] $E_0^{ad}$ ([DFT] $E_0^{ad}$) in examed material systems.

| Properties \ System | | Ni | Cu | $\gamma$-Fe | Pd |
|---|---|---|---|---|---|
| B (GPa) | | 195 | 137 | 281 | 168 |
| $\Omega$ (Å$^3$) | | 2.28 | 2.68 | 2.07 | 2.42 |
| $\Omega_p$ (Å$^3$) | | 2.03 | 2.54 | 1.76 | 2.19 |
| [Model] $E_0^{ad}$ ([DFT] $E_0^{ad}$) (eV) | TET | -2.11 (-2.06) | -1.72 (-1.72) | -1.82 (-1.87) | -2.38 (-2.35) |
| | OCT | -2.24 (-2.26) | -1.86 (-1.85) | -2.14 (-2.18) | -2.40 (-2.38) |
| | BTE | -2.05 (-2.04) | -1.73 (-1.70) | -1.83 (-1.79) | -2.27 (-2.30) |
| | PBP | -1.85 | -1.59 | -1.79 | -2.06 |
| | CTP | -2.20 | -1.75 | -2.04 | -2.18 |

The close agreement between the DFT data and the predictions based on Eq. (4) shown in Fig. 4 demonstrates that Eq. (4) captures the dominant chemomechanical effects of hydrogen adsorption and segregation at GBs. In Eq. (4), $E_0^{ad}$ can be regarded as an intrinsic property of pristine polyhedrons associated with chemisorption of hydrogen, which can be separately determined (other than the GB calculations). The pristine TETs and OCTs are commonly present in bulk FCC lattice and the pristine BTEs are basic constituents of coherent twin boundaries, and the corresponding hydrogen adsorption energies can be readily evaluated. These data, also listed in Table I, are in close agreement with the $E_0^{ad}$ values obtained from the previous GB calculations. This validates the treatment of $E_0^{ad}$ as a material constant. Incidentally, the pristine PBP and OCT polyhedrons are not present in the GB structures examined in the present study. Nonetheless we suggest the possible methods (elaborated in detail in (*28*)) by which one might construct pseudo-pristine PBP and OCT polyhedrons to compute the corresponding values of $E_0^{ad}$. In addition, we note that in Table I, OCT (the polyhedron responsible for H adsorption in bulk lattice) exhibits the lowest $E_0^{ad}$ among all five polyhedrons. Besides $E_0^{ad}$, the partial volume of hydrogen adsorption at GBs, $\Omega_p$, is another important parameter in Eq. (4). Interestingly, the value of $\Omega_p$ obtained is nearly identical to the partial volume of a hydrogen interstitial in the bulk lattice (see Table I). Hence, hydrogen induces similar dilatation both at the GB and in the bulk (*28*).

Eq. (4) provides a predictive model for evaluating the energetics of hydrogen trapping and segregation at GBs. In light of recent experiments (*19, 20, 32, 33*) on hydrogen embrittlement, several case studies of hydrogen embrittlement of GBs were performed, as elaborated in (*28*). Eq. (4) enables a quick evaluation of tendency of hydrogen segregation at GBs. For instance, the $\Sigma 3$ and $\Sigma 3^n$ families (i.e., the "special" GBs defined in (*19*) and discussed earlier in this paper) and $\Sigma 11$ GB exhibit a lack of volume changes of polyhedral structural units. As a result, they are unfavorable to hydrogen trapping and thus less prone to hydrogen embrittlement in terms of less decease of work of separation of GBs. In contrast, other GBs with high sigma numbers, such as $\Sigma 17$ and $\Sigma 73$, involve substantial volume changes of polyhedral structural units and are more susceptible to hydrogen trapping and accumulation, thus giving rise to more severe embrittlement effects (see Figs S8 and S9 in (*28*)). This trend is consistent with the experimental observations of hydrogen embrittlement effects on GB-engineered Ni (*19*) where Ni samples consisting of high density special boundaries (i.e., principally $\Sigma 3$ twin boundaries) are demonstrated to exhibit good HE resistance.

Moreover, our study suggests a mechanistic pathway for further study of the GB effects on hydrogen embrittlement. First, with polyhedrons as atomic structural units of the metal lattice, the diffusion of hydrogen can be considered as discrete hops between neighboring polyhedra. Given the highly localized interactions between hydrogen and GBs, the jumping trial frequency and migration barrier would presumably depend on the coupling of neighboring polyhedra and their associated dilatation. Recognition of such localized interactions will enable the characterization of the complete diffusion parameters through a finite set of calculations (*34*), thus greatly facilitating the study of kinetics of hydrogen migration at GBs. Second, the present study calls for a rigorous continuum micromechanical study on the deformation fields of GBs, e.g., through the generalized Peierls–Nabarro model that treats the atomic interaction right at the GB interface and the continuum elastic interaction for the rest of system (*35, 36*)). This will enable the prediction of volume distortion, $dV_p/V_p^0$ at GBs directly from continuum micromechanics, and thus reduce the need of intensive first principles calculations, besides the intrinsic properties of polyhedrons, such as $E_0^{ad}$ and $\Omega_p$. As such, Eq. (4), in conjunction with the aforementioned analyses, would provide a full-scale predictive framework to quantitatively guide the GB engineering against hydrogen embrittlement.

In summary, we study the energetics of hydrogen adsorption at GBs as a function of structure by combining the space tessellation of polyhedral packing units and the first principles calculations. We further develop a physics-based, predictive model, as given by Eq. (4), to reveal the chemomechanical origin of hydrogen trapping and segregation at GBs. This model is validated through the quantitative evaluation of hydrogen adsorption energies as a function of volumetric deformation of polyhedral structural units at GBs for a variety of FCC metals. Our results advance the atomic-level understanding of the role of GBs in hydrogen embrittlement, and provide mechanistic insights that enable predictive GB engineering against hydrogen embrittlement. Such insight may couple with probability estimates of site occupation by hydrogen to advance quantitative understanding of enhancement of interface decohesion in the presence of hydrogen.

**AKNOWLEDGEMENTS**

J.S. acknowledges the financial support from McGill Engineering Doctoral Award and National Sciences and Engineering Research Council (NSERC) Discovery grant (grant # RGPIN 418469-2012). D.L.M. and T.Z. acknowledge support from QuesTek to study hydrogen effects in metals. Z.X. acknowledges the financial support from China Scholarship Council (CSC). We also acknowledge Supercomputer Consortium Laval UQAM McGill and Eastern Quebec for providing computing power.

**Supplementary Material for Chemomechanical Origin of Hydrogen Trapping at Grain Boundaries in FCC Metals**

Xiao Zhou[a], Daniel Marchand[a], David L. McDowell[b], Ting Zhu[b], Jun Song[a, *b]

[a]Department of Mining and Materials Engineering, McGill University, Montréal, Québec H3A 0C5, Canada

[b]Woodruff School of Mechanical Engineering, Georgia Institute of Technology, Atlanta, Georgia 30332, USA

## S1. COMPUTATIONAL METHODOLOGY

First-principles density functional theory (DFT) calculations were performed using the Vienna *ab initio* Simulation Package (VASP) (*1, 2*) on the basis of the density functional theory (DFT) with the projector augmented wave method (PAW) (*3, 4*). The generalized gradient approximation (GGA) (*5*) with the Perdew-Burke-Ernzerhof exchange-correlation (PBE) functional (*6*), and the first-order Methfessel-Paxton scheme with a smearing of 0.05 eV were implemented in the calculations. A k-point grid of 3×1×5 and a cut-off energy of 400 eV were used and shown to yield sufficient accuracy. Structural relaxation was considered converged when atomic force was less than 0.01 eV/Å. During relaxation, atoms were allowed to move freely except for the atoms in the surface layers (i.e., see Fig. 1) whose motions were constrained to be perpendicular to the GB only. A vacuum space of 10 Å in thickness exists between surfaces to eliminate interactions between surfaces. A variety of GBs were examined, with most of them shown in Fig. S1.

---

*[b] Author to whom correspondence should be addressed. E-Mail: jun.song2@mcgill.ca

We focused on symmetric tilt GBs in the present study, yet several non-symmetric GBs were also investigated for the purpose of *proof-of-concept* or when looking for certain polyhedrons.

In our calculations, the zero point energy (ZPE) correction was not included due to the high computational cost. However, preliminary studies with ZPE correction were performed, showing that ZPE does not affect our results (*albeit* possibly overall a small shift for energetics that does not change the trend).

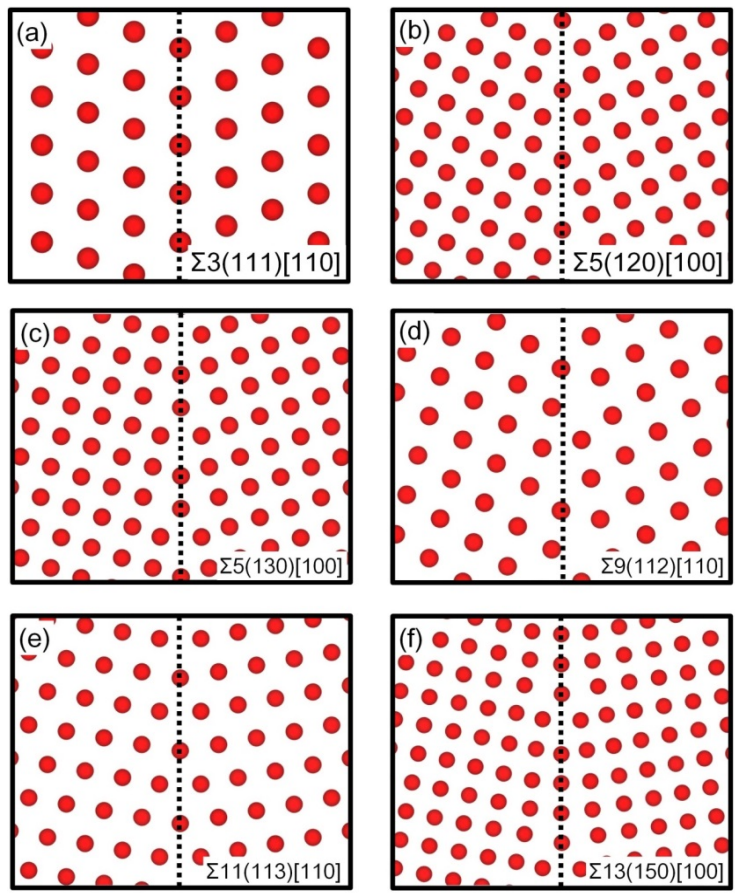

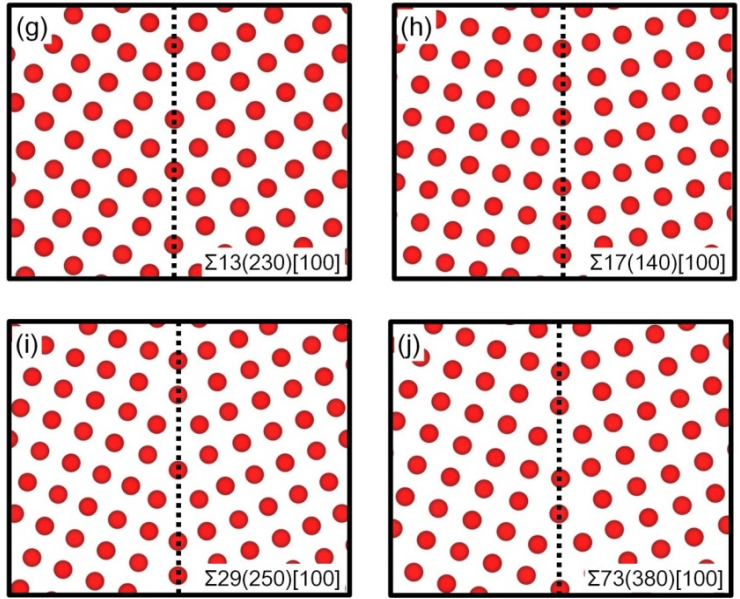

**Figure S1**: The atomistic configurations of the symmetric tilt GB structures studied.

## S2. SPACE TESSELLATION ALGORITHM FOR QUICK IDENTIFICATION OF POLYHEDRONS

A space tessellation algorithm based on 3D Voronoi diagram (*7*) was developed to enable quick division of the lattice into convex deltahedron units. The algorithm first computes the locations of all the vertex points of the Voronoi polygons enclosing the host atoms. For each vertex point, the host atoms that neighbor it (within a cut-off distance) are identified. Then with the vertex point approximated as the polyhedron center, Voronoi diagrams are constructed to determine possible polyhedron configurations. The above process is illustrated in Fig. S2. One main challenge in the above algorithm is that there are degenerated Voronoi cell vertices (*8*), i.e., multiple vertex points being in close vicinity of each other (the red dash circle, as shown in Fig. S2 (a)). Herein, we introduce a simple method to consolidate degenerated points, i.e., all vertex points with mutual separations within a certain cut-off distance are "*fused*" into a single point (e.g., a cut-off distance of 1.30 Å is chosen for the Ni system). This approach is schematically illustrated in Fig. S2b. The central idea behind the above tessellation algorithm is that the vertices of Voronoi cells, after consolidation to eliminate degeneracy, effectively serve as the vertices of polyhedrons.

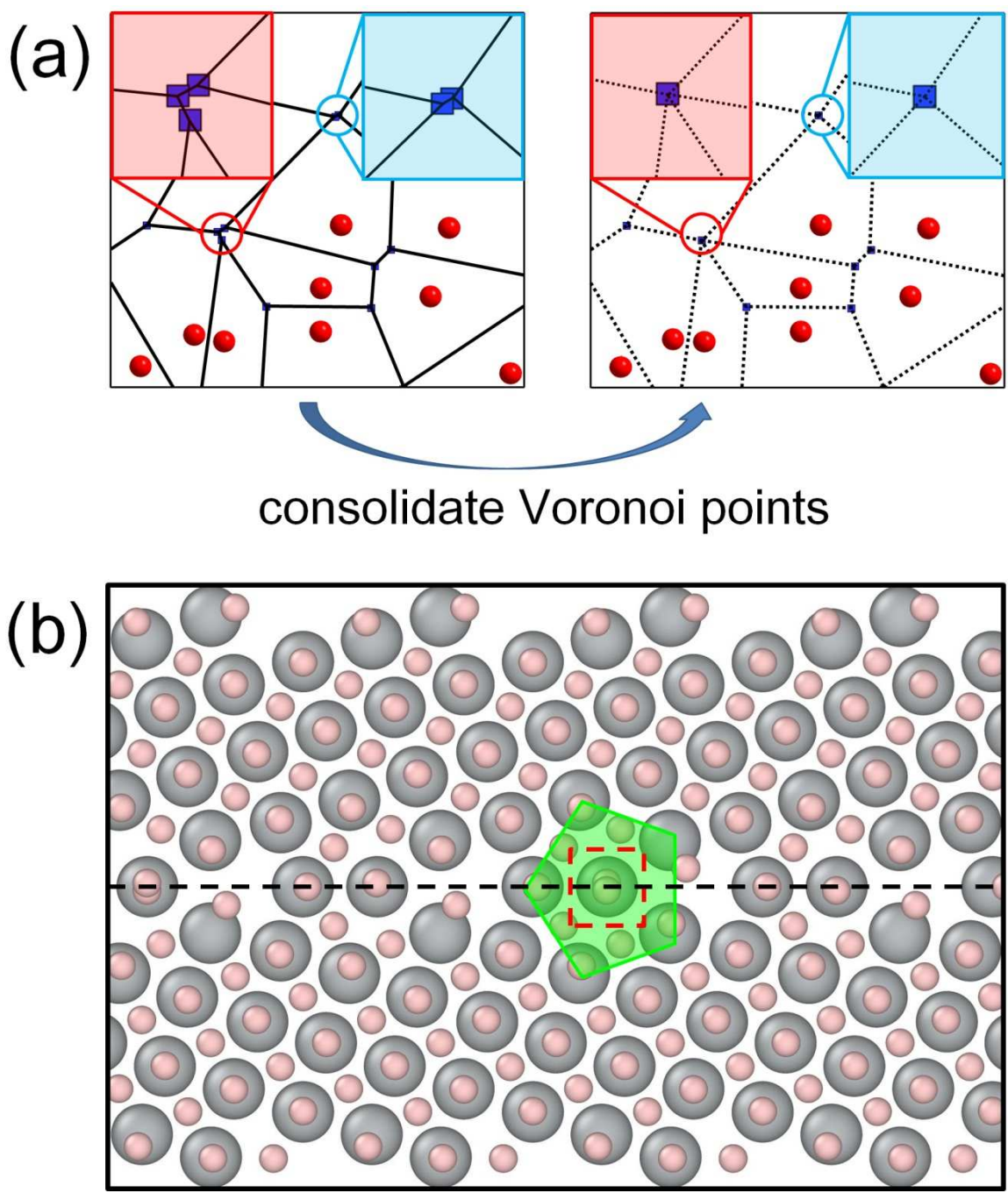

**Figure S2**: (a) A schematic illustration of the consolidation process of degenerated Voronoi vertices. The subfigure on the left shows a raw Voronoi tessellation plot where the red spheres indicate seed points, the small black squares and solid black lines indicate the resultant vertices and line segments. The red and cyan circles indicate two sample regions where degeneration of Voronoi vertices occur, with the red-shaded and cyan-shade squares respectively showing their zoom-in views. The subfigure on the right shows the plot after the consolidation process. (b) A projection view of $\Sigma 5(130)[100]$ GB where the large grey spheres represent host atoms and the pink balls represents the Voronoi vertices resulted from space tessellation. The red dashed square encloses a region where the degeneration of Voronoi vertices occurs. Those degenerated vertices within the red dashed square are subsequently "fused" into a single point following the process outlined in (a). The green-shaded polygon indicates a pentagonal bipyramid (PBP) from the space tessellation algorithm (see details in Section S2).

## S3. MATERIAL PROPERTIES OF BULK LATTICES

Prior to GB calculations, DFT calculations were performed to obtain the essential material properties of the bulk lattice. A 3×3×3 periodic supercell containing 108 host atoms was used in these calculations. Table S1 lists the values of equilibrium lattice constant ($a_0$), bulk modulus ($B$) and partial volume ($\Omega$), for different materials (i.e., Ni, Cu, FCC-Fe and Pd). Those values are in good agreement with data in literatures, as shown in Table S1.

**Table S1**: Essential material properties, i.e., equilibrium lattice constant ($a_0$), bulk modulus ($B$) and partial volume ($\Omega$) for Ni, Cu, $\gamma$-Fe and Pd.

|  | Ni | $\gamma$-Fe | Cu | Pd |
|---|---|---|---|---|
| $a_0$ (Å) | 3.52 | 3.46 | 3.64 | 3.95 |
| Ref. | 3.52[9];3.53[10];3.38[11];3.47[12]. | 3.45[13] | 3.65[10];3.47[11];3.56[12] | 3.96[10];3.79[11];3.83[12] |
| $B$(GPa) | 195 | 281 | 137 | 168 |
| Ref. | 186[14];221[10];265[11];192[15] | 314[16] | 117[10];192[11];137[12] | 171[10];232[11];181[12] |
| $\Omega$ (Å$^3$) | 2.28 | 2.07 | 2.68 | 2.42 |
| Ref. | 2.20[17]; 3.30[18]; 2.30[15] | - | - | 2.47[17] |

## S4. PREDICTIVE MODELING OF HYDROGEN ADSORPTION ENERGETICS

## S4.1. *Localization of hydrogen-lattice interactions*

As mentioned in the main text, the hydrogen-lattice interactions are strictly local and thus the adsorption energy is solely determined by the local environment of the polyhedron that encloses the hydrogen atom. The localization was demonstrated via the differential charge density plots in Fig. 2 for Ni. Similarly for other material systems, i.e., Cu, $\gamma$-Fe and Pd, the localization was also observed, illustrated in Fig. S3 below.

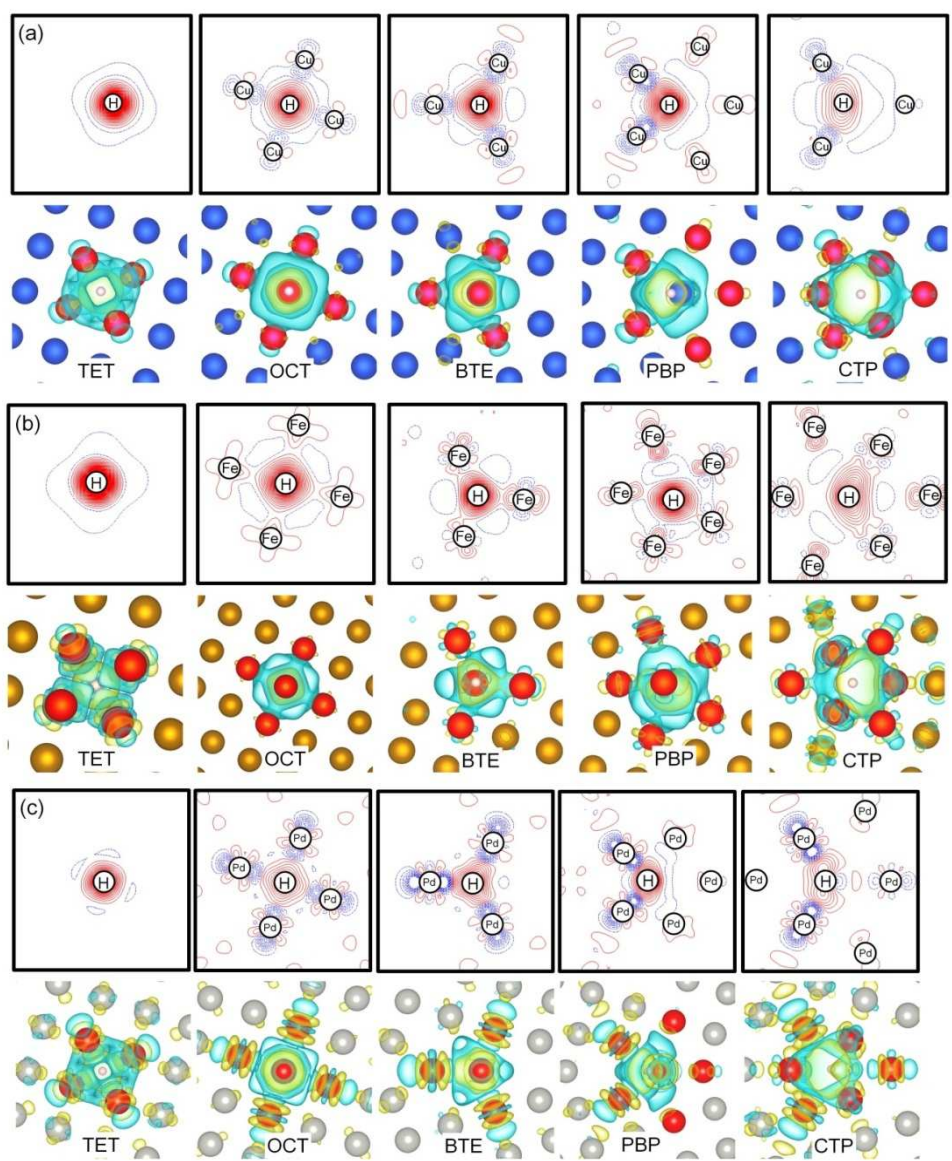

**Figure S3**: 2D and 3D differential charge density plots of Σ 5(130)[100] GB projected on the (100) plane for (a) Cu, (b) γ-Fe, and (c) Pd. The blue dash and red solid lines in the 2D plots represent electron depletion and electron accumulation respectively. In the 3D plots, blue isosurface and yellow isosurface are electron depletion and electron accumulation, respectively. The small pink spheres are hydrogen atoms and big red spheres are metal atoms at polyhedron vertices. The big gray spheres represent other metal atoms surrounding the polyhedron.

**S4.2. *Determination of model parameters***

Applying the proposed analytical model to the $E^{ad}$ vs $dV_p/V_p^0$ data for all polyhedrons, the model parameters $\Omega_p$ and $E_0^{ad}$ can be determined. The global least-squares method (*19*) was employed, and from respectively the slope and intercept values $\Omega_p$ and $E_0^{ad}$ can be determined. Meanwhile as mentioned in the main text, $E_0^{ad}$ is an intrinsic property of the pristine polyhedron and thus can be determined separately, while can then be used to validate the analytical model. The validation can readily performed for polyhedrons TET, OCT and BTE, the pristine forms of which are available in bulk lattice and we have demonstrated in the main text that the two sets of data (i.e., the ones from fitting and the ones separately computed) show excellent agreement. On the other hand, the pristine PBP and CTP polyhedrons are not available (at least in all the GB structures examined in the present study).

In our preliminary pursuit to tackle this challenge, we found that one possible route to construct *pseudo*-pristine PBP and CTP polyhedrons, using which the corresponding $E_0^{ad}$ values can be determined. Below we outline two potential means attempted for the construction of those *pseudo*-pristine polyhedrons.

1) **Method #1**: A standing-alone pristine polyhedron with atoms sitting at vertices

and the edge length being $\sqrt{2}a_0/2$ is built, following which extra metal atoms are introduced to add in the region surrounding the polyhedron. In particular, one atom will be added one top of each polyhedron surface. These extra atoms form a "shell" enclosing the polyhedron in the core. The resultant configuration is then optimized with a minimal interatomic separation of $\sqrt{2}a_0/2$ enforced[c]. The afore-mentioned shell construction may be repeated to build several shells. A hydrogen atom is then introduced into the polyhedron and the whole system is then relaxed (with atoms in the outmost shell fixed while other atoms free to move) to obtain the hydrogen adsorption energy (as the value for $E_0^{ad}$). Two sample constructions (for PBP and CTP) are illustrated in Fig. S4. The number of shells and the details of how each shell is constructed depend on the convergence and polyhedron stability during the relaxation process;

2) **Method #2**: Given than distorted PBP and CTP exist along certain GBs, another way to construct the pseudo-pristine polyhedron is to carve out a region with the polyhedron of interest in the center. A displacement field can then be applied to the carved-out cluster to "revert" the distortion so as to transform the polyhedron into its pristine form. Then a hydrogen atom can be introduced to compute the hydrogen adsorption energy (as the value for $E_0^{ad}$) – again with the outmost atoms fixed while other atoms free to move. For this method, the challenge is to properly determine the size of the carve-out region and systematically define the displacement field.

---

[c] During the process, the atoms belonging to the polyhedron are held fixed.

We can see that the above two methods of *pseudo*-pristine polyhedron construction are rather *ad-hoc* in nature, which definitely necessitates detailed studies to grant a consistent and reliable construction - this will be tackled in our future studies.

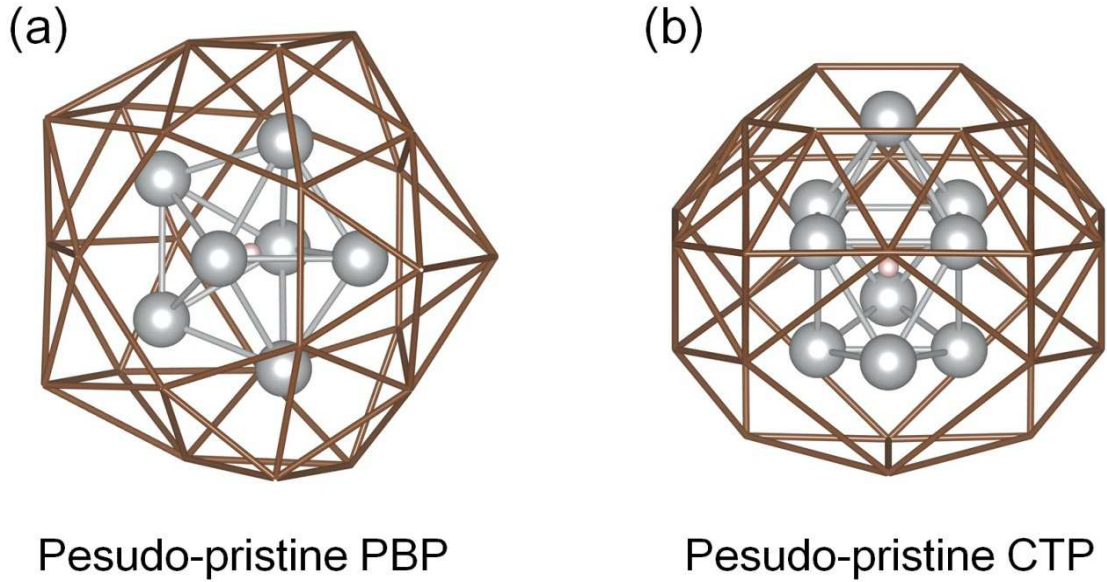

**Figure S4**: Sample construction of *pseudo*-pristine (a) PBP and (b) CTP polyhedron configurations.  The grey spheres represent atoms at the pristine polyhedron vertices while the small pink spheres represent the hydrogen atom enclosed in the polyhedron; while the brown bonds denote the bonds between atoms in the outer shells surrounding the polyhedron.

The $E_0^{ad}$ data obtained using the *pseudo*-pristine polyhedrons are shown in Fig. S5, in comparison with the data obtained from fitting (see Table I in main text) from which we can see that despite being somewhat scattered they are in reasonable agreement with the values obtained from GB simulations.   One exception is the PBP polyhedron in Pd, where large deviation between $E_0^{ad}$ values (from fitting and the *pseudo*-pristine approach) can be observed.   This discrepancy may originate from the fact that Pb has the largest lattice constant (3.954 Å), which results in very large volumes for PBP and CTP, then prevent the hydrogen atom from settling in an optimum state when a non-periodic cluster model is used in the DFT calculation.

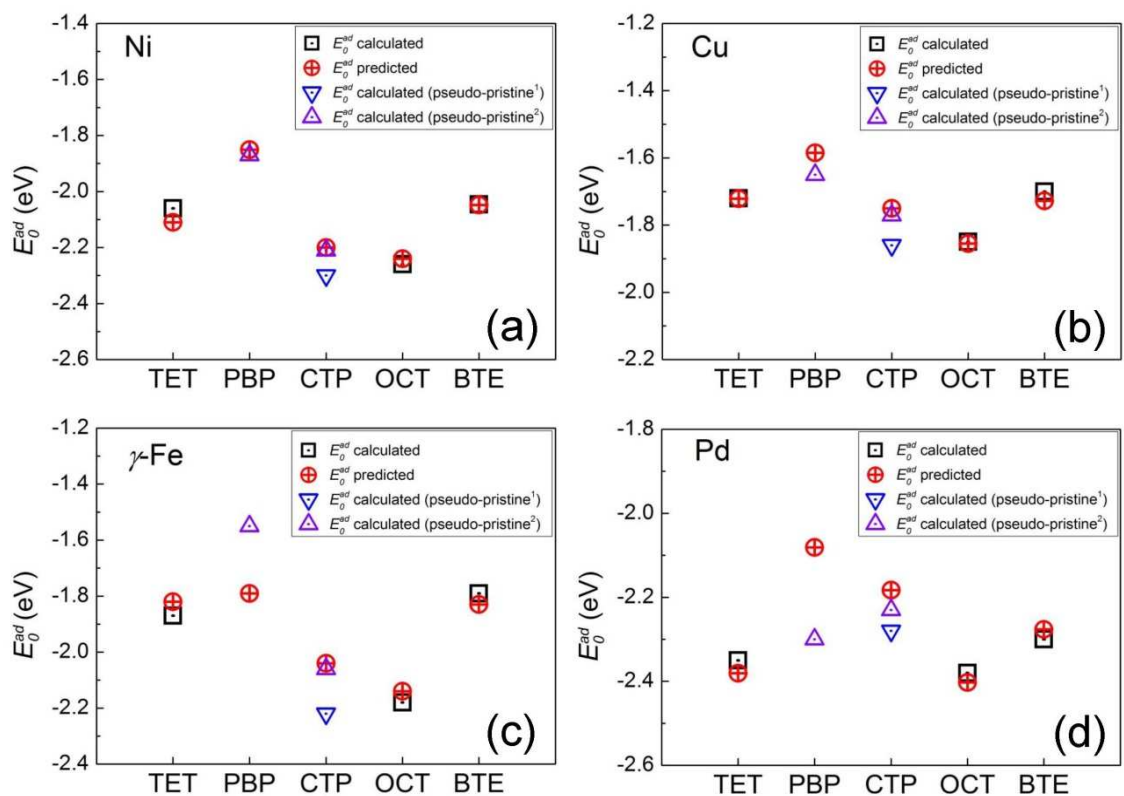

**Figure S5**: Comparison between the predicted $E_0^{ad}$ values obtained from the GB data (i.e., $E_0^{ad}$ predicted, red open circles), predicted $E_0^{ad}$ values for pristine polyhedrons in bulk lattice (i.e., $E_0^{ad}$ calculated, black solid squares, for TET, OCT and BTE polyhedrons), and pristine polyhedrons using the pseudo-polyhedron construction (i.e., $E_0^{ad}$ calculated [pseudo-pristine$^n$], for PBP and CTP polyhedrons, with $n$ = 1 or 2 indicated the construction methods #1 or #2 respectively), for (a) Ni, (b) Cu, (c) $\gamma$-Fe and (d) Pd.

## S5. HYDROGEN SEGREGATION ANALYSIS

The segregation energy of hydrogen at GB and free surface is calculated as:

$$E_{g(fs)}^{sg} = E_{g(fs)}^{ad} - E_{b}^{ad}, \tag{S1}$$

where $E_{g(fs)}^{ad}$ denotes the adsorption energy of hydrogen at GB (free surface), and $E_{b}^{ad}$ is the adsorption energy of hydrogen in bulk lattice.

The hydrogen occupancy $c_i$ (defined as hydrogen/metal atomic ratio) at an adsorption site $i$ can be related to the bulk hydrogen concentration $c_0$ as (*20, 21*):

$$\frac{c_i}{1-c_i} = \frac{c_0}{1-c_0} \exp(-E_i^{sg}), \tag{S2}$$

where the segregation energy of hydrogen at site $i$ is defined in the same way as Eq. S1. Combining the above with the analytical model (see Eq. 4 in the main text), we can obtain the segregation energy for an interstitial site enclosed in a polyhedron as:

$$E_i^{sg} = -B\Omega_p \frac{dV_p^i}{V_p^{0,i}} + E_{0,i}^{ad} - E_b^{ad}, \tag{S3}$$

where $dV_p^i / V_p^{0,i}$ and $E_{0,i}^{ad}$ denote the corresponding $dV_p / V_p^0$ and $E_0^{ad}$ (cf. Eq. 4 in the main text) of the polyhedron at adsorption site $i$. From Eqs S2-S3, the hydrogen occupancy at different polyhedrons as functions of $dV_p / V_p^0$ and $c_0$ can be analyzed, shown in Fig. S6 (for the sample case of T=300K).

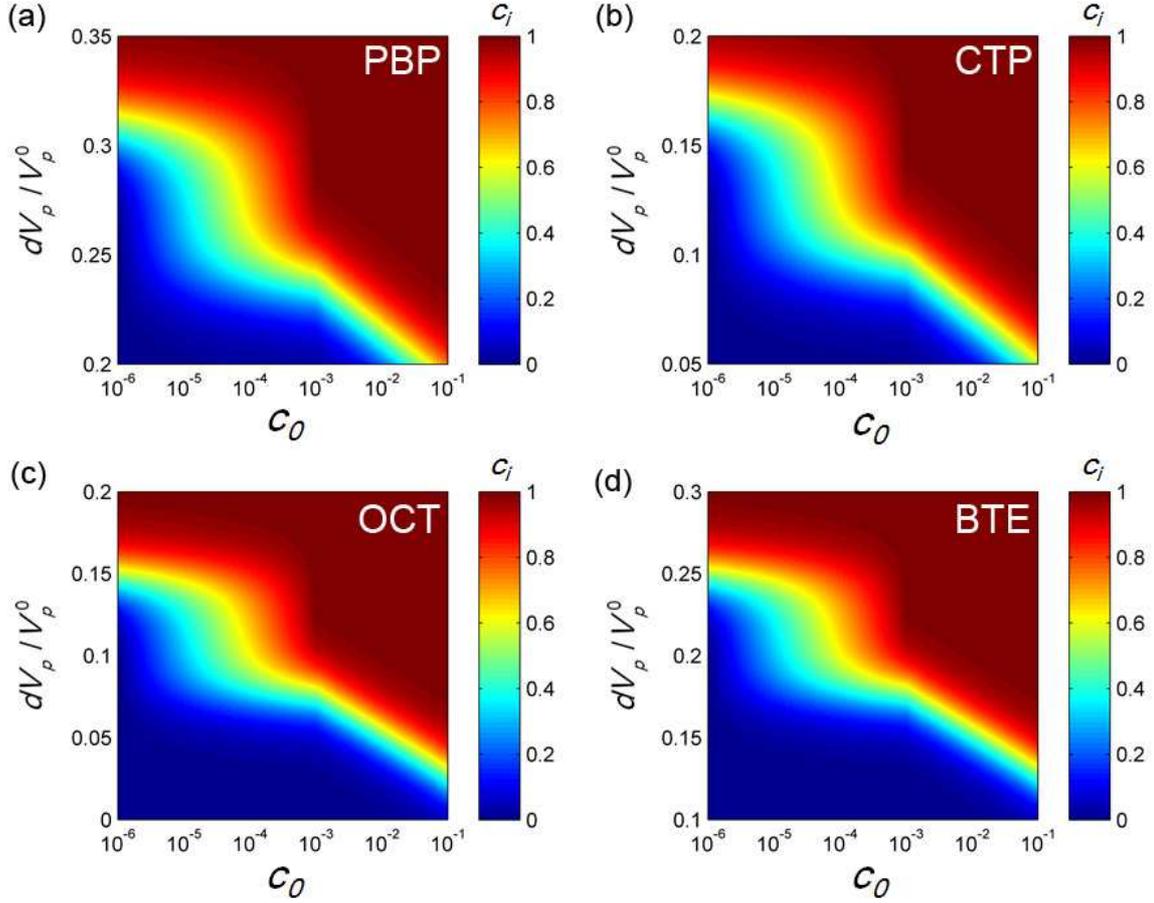

**Figure S6:** Hydrogen occupancy $c_i$ at different polyhedrons as the volume distortion $dV_p/V_p^0$ and bulk hydrogen concentration $c_0$ vary.

With hydrogen segregation at GBs characterized, the hydrogen influenced work of separation can be easily analyzed combining our analytical model and the thermodynamic Rice-Wang model (22): $W_c \approx W_0 - \sum_i (E_{gb}^{sg,i} - E_{fs}^{sg,i}) c_i \Theta_{max}$, where $W_c$ is the work of separation with hydrogen trapping, $W_0$ is the work of separation without any hydrogen, $\Theta_{max}$ is the maximum hydrogen concentration of polyhedral site. Fig. S7 shows the predicted evolution of work of separation ($W_c$) for several representative symmetric tilt Ni GBs, i.e., Σ 3, Σ 9, Σ 11 Σ 17 and Σ 73 GBs. Here only the hydrogen segregation directly along the GB was considered in the analysis for simplicity and the room temperature of $T$=300K was assumed.

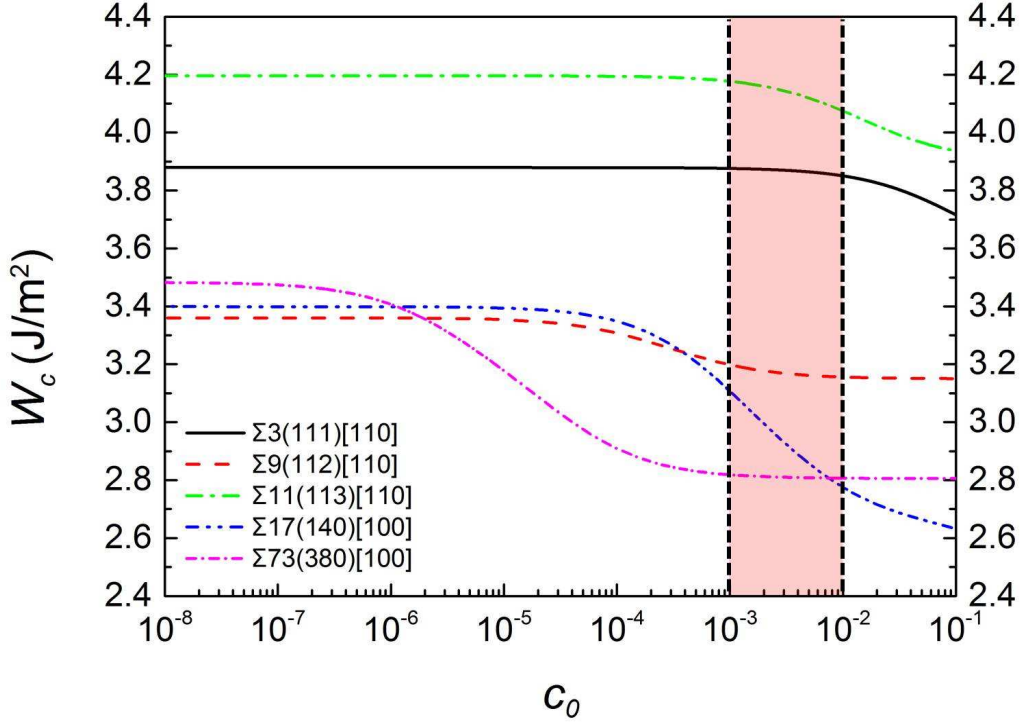

**Figure S7**: The evolution of work of separation ($W_c$) for several symmetric tilt Ni GBs, i.e., $\Sigma 3(111)[110]$ (solid black line), $\Sigma 9(112)[110]$ (red dash line), $\Sigma 11(113)[110]$ (green dash dot line), $\Sigma 17(140)[100]$ (blue dash dot-dot dash line), $\Sigma 73(380)[100]$ (pink short dash dot line) as the hydrogen bulk concentration varies (considering hydrogen segregation along the GB).

There are several observations from Fig. S7. First, we see that $\Sigma 73$ is most sensitive to H infiltration, showing an apparent drop in $W_c$ starting from $c_0$ = 1 appm, while $\Sigma 3$ is most resistant to H infiltration with $W_c$ remaining invariant even up to a very high $c_0$ of 10,000 appm. Also it is worth noting from Fig. S7 that another GB, $\Sigma 11$, that exhibits very similar behaviors as $\Sigma 3$. Second, we note that the two GBs, $\Sigma 17$ and $\Sigma 73$, exhibit noticeably larger and more abrupt drop in $W_c$ in comparison to the other three, i.e., $\Sigma 3$, $\Sigma 9$ and $\Sigma 11$, suggesting that $\Sigma 17$ and $\Sigma 73$ be more susceptible to HE (in the sense of cleavage-type failure).

To put the above observations in perspective, we examined relevant experimental studies reported in literature. In a recent study by S. Bechtle et al. (*23*), the HE phenomena in pure Ni was investigated. In the study, they demonstrated that the HE resistance of Ni can be considerably enhanced by increasing the fraction of the so-called "special" GBs. These "special" GBs refer to the GBs with low sigma number ($\Sigma<29$) under the CSL nomenclature (*24*). Among those "special" GBs, $\Sigma 3$ and $\Sigma 3^n$ GBs were shown to play the main role in enhancing the HE resistance (*23*). This is directly reflected in our results shown in Fig. S7, where we see that $\Sigma 3$ and $\Sigma 9$ are both very resistant to hydrogen infiltration. On the other hand, our results also indicated that the general definition of "special" GBs is ambiguous, as evidenced by the case of $\Sigma 17(140)[100]$ (a "special" GB according to the general definition) in Fig. S7 where sizable influence of hydrogen is observed. This echoes with the work by Randle (*24*) which demonstrated that the CSL definition of "special" GBs is indeed ambiguous and a more realistic definition is needed when applied to corrosion or fracture of GBs or GB engineering).

In light of the above, we explored the possibility of identifying a more-refined definition of "special" GBs using the new insights from the present study. As we stated in the main text, the key underlying strong trapping of hydrogen at a GB is the lattice distortion. In Fig. S8, we show the corresponding contour plots of lattice distortion $dV_p/V_p^0$ (of polyhedrons) along those "special" ($\Sigma<29$) GBs examined in Fig. S7, we note that these contour plots also help effectively distinguish the GBs within the "special" category, i.e., we can clearly see that $\Sigma 3$, $\Sigma 9$ and $\Sigma 11$ show significantly less degree of lattice distortion than $\Sigma 17$. In this regard, we expect that $\Sigma 17$ would remain quite susceptible to HE despite it being a "special" GB (according to the general $\Sigma <29$ definition). Among the three GBs, i.e., $\Sigma 3$, $\Sigma 9$ and $\Sigma 11$, we see from Figs S7 and S8 that $\Sigma 3$ and $\Sigma 11$ are particularly insensitive to hydrogen infiltration. This can be further explained by a detailed analysis at the polyhedron level, described as follows. For the $\Sigma 3$ GB, there is simply no distortion at GB as it consists of only pristine BTE polyhedrons. While for the $\Sigma 11$ GB, it is composed of nearly pristine CTP polyhedrons with ignorable distortion. The observed high resistance of $\Sigma 11$ to hydrogen is also consistent with previous experimental studies, e.g., Baskes *et al* (*25*) investigated H trapping behavior in Ni $\Sigma 9$ and $\Sigma 11$ GB, suggesting no apparent HE tendency; Du *et al.* (*13*) demonstrated there is no notable hydrogen trapping in $\gamma$-Fe $\Sigma 11$ GB. From the above, we see that we can redefine "special" GBs in the context of HE, e.g., as the group of GBs with overall (or individual) lattice distortion smaller than a *critical threshold*. Of course further studies are necessary to properly determine this critical threshold distortion.

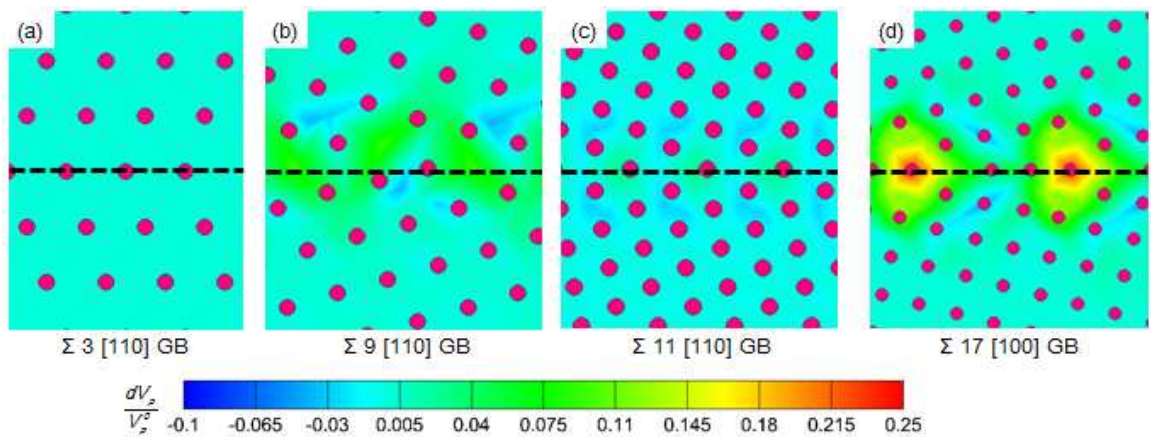

**Figure S8**: Contour plots of lattice distortion $dV_p/V_p^0$ (of polyhedrons) along and neighboring the GB for (a) Σ3, (b) Σ9, (c) Σ11 and (d) Σ17 symmetric tilt GBs in Ni.

# SUPPLEMENTARY REFERENCES